\documentclass{article}

\usepackage{microtype}
\usepackage{graphicx}
\usepackage{subfigure}
\usepackage{booktabs} 

\usepackage{hyperref}

\usepackage[accepted]{ml4astro2025}

\usepackage{amsmath}
\usepackage{amssymb}
\usepackage{mathtools}
\usepackage{amsthm}

\usepackage[capitalize,noabbrev]{cleveref}

\theoremstyle{plain}

\theoremstyle{definition}

\theoremstyle{remark}

\usepackage[textsize=tiny]{todonotes}

\mlforastrotitlerunning{Image-Based Multi-Survey Classification of Light Curves with a Pre-Trained Vision Transformer}

\begin{document}

\twocolumn[
\mlforastrotitle{Image-Based Multi-Survey Classification of Light Curves with a Pre-Trained Vision Transformer}



\mlforastrosetsymbol{equal}{*}

\begin{mlforastroauthorlist}
\mlforastroauthor{Daniel Moreno-Cartagena}{udec}
\mlforastroauthor{Guillermo Cabrera-Vives}{udec,cdia,mas,eee}
\mlforastroauthor{Alejandra M. Muñoz Arancibia}{mas,cmm}
\mlforastroauthor{Pavlos Protopapas}{harvard}
\mlforastroauthor{Francisco Förster}{idia,mas,cmm}
\mlforastroauthor{Márcio Catelan}{iauc,mas,caiuc}
\mlforastroauthor{A. Bayo}{eso}
\mlforastroauthor{Pablo A. Estévez}{eeudechile,mas}
\mlforastroauthor{P. Sánchez-Sáez}{eso}
\mlforastroauthor{Franz E. Bauer}{tarapaca}
\mlforastroauthor{M. Pavez-Herrera}{iauc}
\mlforastroauthor{L. Hernández-García}{titans,mas,ifavalpo}
\mlforastroauthor{Gonzalo Rojas}{udec,cdia}
\end{mlforastroauthorlist}

\mlforastroaffiliation{udec}{Department of Computer Science, Universidad de Concepción, Edmundo Larenas 219, Concepción, Chile}
\mlforastroaffiliation{cdia}{Center for Data and Artificial Intelligence, Universidad de Concepción, Edmundo Larenas 310, Concepción, Chile}
\mlforastroaffiliation{mas}{Millennium Institute of Astrophysics (MAS), Santiago, Chile}
\mlforastroaffiliation{eee}{Heidelberg Institute for Theoretical Studies, Heidelberg, Germany}
\mlforastroaffiliation{cmm}{Center for Mathematical Modeling (CMM), Universidad de Chile, Santiago, Chile}
\mlforastroaffiliation{harvard}{John A. Paulson School of Engineering and Applied Sciences, Harvard University, Boston, MA, USA}
\mlforastroaffiliation{eso}{European Southern Observatory (ESO), Karl-Schwarzschild-Strasse 2, 85748 Garching bei München, Germany}
\mlforastroaffiliation{tarapaca}{Instituto de Alta Investigaci{\'{o}}n, Universidad de Tarapac{\'{a}}, Casilla 7D, Arica, Chile}
\mlforastroaffiliation{idia}{Data and Artificial Intelligence Initiative (IDIA), Faculty of Physical and Mathematical Sciences, Universidad de Chile}
\mlforastroaffiliation{iauc}{Instituto de Astrofísica, Pontificia Universidad Católica de Chile, Av. Vicuña Mackenna 4860, 7820436 Macul, Santiago, Chile}
\mlforastroaffiliation{caiuc}{Centro de Astro-Ingeniería, Pontificia Universidad Católica de Chile, Av. Vicuña Mackenna 4860, 7820436 Macul, Santiago, Chile}
\mlforastroaffiliation{titans}{Millennium Nucleus on Transversal Research and Technology to Explore Supermassive Black Holes (TITANS), Gran Breta\~na 1111, Playa Ancha, Valpara\'iso, Chile}
\mlforastroaffiliation{ifavalpo}{Instituto de F\'isica y Astronom\'ia, Facultad de Ciencias, Universidad de Valpara\'iso, Gran Breta\~na 1111, Playa Ancha, Valpara\'iso, Chile}
\mlforastroaffiliation{eeudechile}{Dept. of Electrical Engineering, University of Chile, Santiago, Chile}

\mlforastrocorrespondingauthor{Daniel Moreno-Cartagena}{dmoreno2016@inf.udec.cl}

\mlforastrokeywords{Machine Learning, ICML}

\vskip 0.3in
]



\printAffiliationsAndNotice{}  

\begin{abstract}
We explore the use of \textsc{Swin Transformer V2}, a pre-trained vision Transformer, for photometric classification in a multi-survey setting by leveraging light curves from the Zwicky Transient Facility (ZTF) and the Asteroid Terrestrial-impact Last Alert System (ATLAS). We evaluate different strategies for integrating data from these surveys and find that a multi-survey architecture which processes them jointly achieves the best performance. These results highlight the importance of modeling survey-specific characteristics and cross-survey interactions, and provide guidance for building scalable classifiers for future time-domain astronomy.
\end{abstract}

\section{Introduction}
\label{sec:introduction}

The field of time-domain astronomy is entering a new era of data abundance and interoperability. Over the past two decades, large-scale sky surveys with diverse cadences, spectral coverage, and sky footprints have expanded the observational landscape \citep{Catelan}. Surveys such as the Zwicky Transient Facility \citep[ZTF;][]{bellm2018zwicky}, Asteroid Terrestrial-impact Last Alert System \citep[ATLAS;][]{tonry2018atlas}, Panoramic Survey Telescope and Rapid Response System \citep[Pan-STARRS1;][]{kaiser2010pan}, {\em Gaia} \citep{gaia2023gaia}, and Optical Gravitational Lensing Experiment \citep[OGLE;][]{Udalski2015}, among others \citep{drake2012catalina, shappee2014man, Catelan}, have independently cataloged millions of celestial sources, providing a rich yet heterogeneous view of variability across the universe. In this context, photometric classification, the task of identifying and categorizing variable and transient objects based on their light curves, has become a central challenge in modern time-domain astronomy. 

Historically, most machine learning approaches to photometric classification have relied on training models with light curves from a single survey \citep{muthukrishna2019rapid, villar2019supernova, jamal2020neural, sanchez2021alert, allam2024paying, becker2025}, often using data augmentation and feature extraction to improve predictions within this setting \citep{lochner2016photometric, boone2019avocado, gomez2020classifying, villar2020superraenn, garcia2022improving, monsalves2024application}. More recently, transfer learning has been employed to adapt models trained on one survey to another \citep{mahabal2017deep, carrasco2019deep, pimentel2022deep, gupta2025transfer}, and self-supervised learning has been used to produce survey-agnostic representations that can generalize across different surveys \citep{pan2022astroconformer, donoso2023astromer, moreno2023positional, cadiz2024workshop, cadiz2024Journal}. However, these methods still rely on models trained on light curves from individual surveys, without jointly processing data from multiple telescopes. In parallel, multimodal approaches that combine light curves with other data modalities (e.g., spectra, metadata, derived features) \citep{cabrera2024atat, rizhko2024astrom, rizhko2024self, zhang2024maven} have shown promise, highlighting their potential to improve predictive performance, but they still operate within single-survey settings. While several works have begun exploring multi-survey frameworks \citep{RichardsMS, aguirre2019deep, murrayunsupervised, engel2024preliminary}, explicit methods for jointly exploiting complementary information across surveys remain an open area of research.

In this work, we build upon the straightforward approach proposed by \citet{moreno2025leveraging}, who fine-tuned a \textsc{Swin Transformer V2} (\textsc{SwinV2}), a hierarchical vision Transformer pre-trained on \texttt{ImageNet-21K} \citep{liu2022swin}, for multi-band light curve classification. We extend this approach to the multi-survey setting by exploring fusion strategies to combine data from ZTF and ATLAS, two complementary surveys with overlapping sky coverage and distinct observational characteristics. We propose an architecture that jointly models light curves from different telescopes, using the vision Transformer as its core. To enable this, light curves are transformed into image-based representations that serve as input to the model. Our results suggest that incorporating multiple surveys can improve classification performance, that simple ensemble approaches already leverage complementary information, and that naive early fusion underperforms relative to single-survey baselines. Moreover, ATLAS contributes complementary information despite lower individual performance.

\section{Methods}
\label{sec:methods}

\subsection{Multi-Band Light Curves to Images}

\begin{figure}[t]
\vskip 0.2in
\begin{center}
\centerline{\includegraphics[width=8cm]{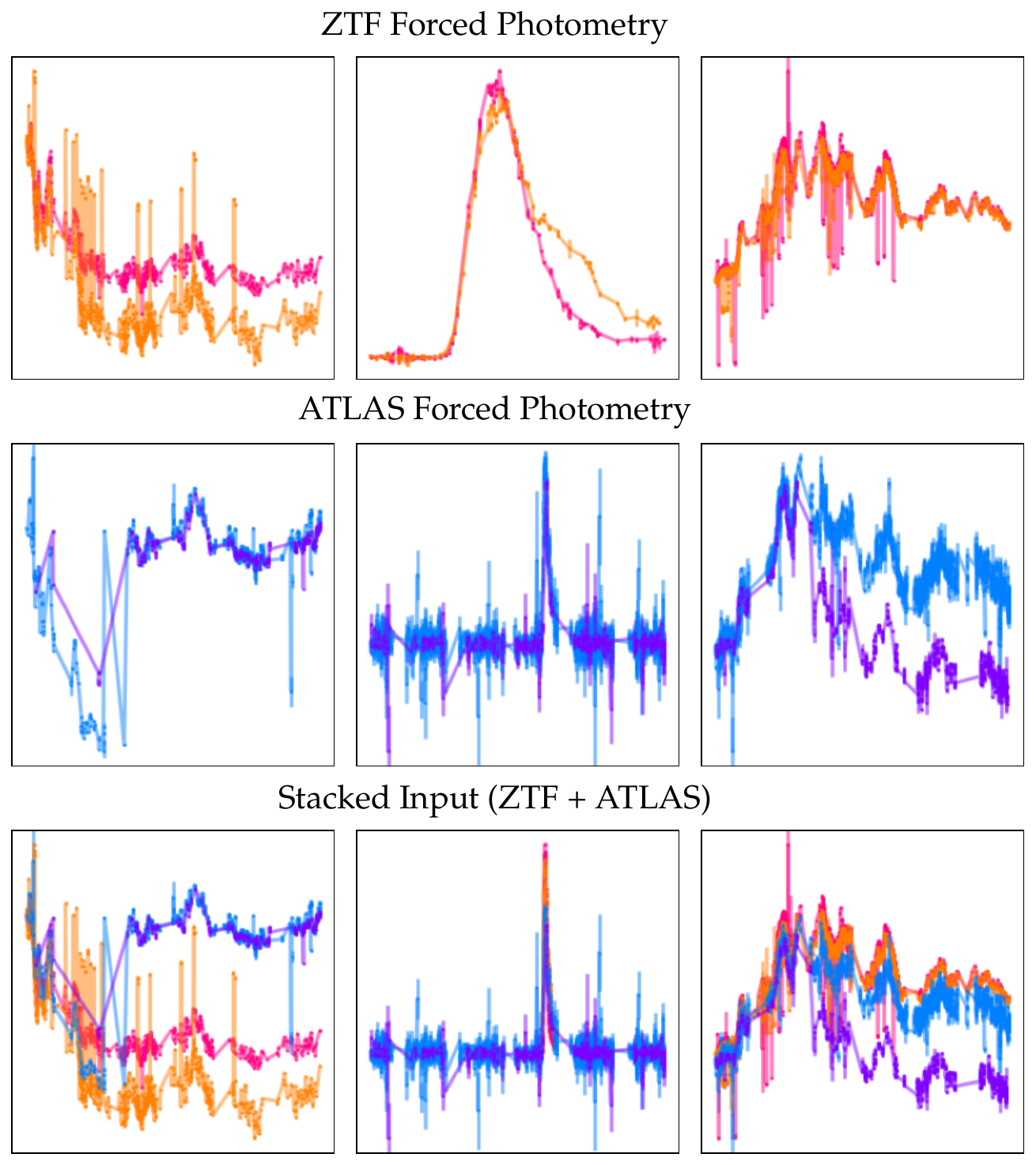}}
\caption{Model input images constructed from forced photometry light curves. Rows show ZTF ($g$ and $r$ bands), ATLAS ($c$ and $o$ bands), and the stacked input with all bands. Images represent flux difference over time, with axes omitted since time and flux are not explicitly used. From left to right, columns show ZTF18aajpcoq (Blazar), ZTF21abfnfdw (SNIa), and ZTF18aampkkf (AGN).}
\label{fig:input_model}
\end{center}
\vskip -0.4in
\end{figure}

\label{sec:lc_to_images}
In our approach, multi-band light curves are transformed into images used as input to a pre-trained vision Transformer. We adopt the Overlay visualization method proposed by \citet{moreno2025leveraging}, in which observations from different photometric bands are plotted simultaneously within a unified visual space, as illustrated in Figure~\ref{fig:input_model}. To ensure comparability across light curves, we apply Min-Max normalization independently to both time and flux values for each light curve, jointly considering all its bands. This normalization maps all values to the range $[0, 1]$, removing information about absolute time and brightness and encouraging the model to focus on the shape of the light curve rather than its amplitude or temporal scale to discriminate between classes. Global normalization across the dataset was explicitly avoided, as it disproportionately suppressed low-amplitude variability and negatively impacted classification performance. Additionally, flux errors are scaled using the same Min-Max factor applied to the flux, preserving relative uncertainty information. Figure~\ref{fig:input_model} shows three example input images per dataset, including lines, points, and error bars (see Section~\ref{sec:training_details}). Each band is mapped to a location near a primary RGB axis, ensuring balanced color representation and avoiding visual bias across bands.

\subsection{Vision Transformer Model}
\label{sec:vt_model}
Following the methodology introduced by \citet{moreno2025leveraging}, we adopt the pre-trained \textsc{SwinV2} architecture, adapted for the classification of light curves, as the core model. \textsc{SwinV2} is a hierarchical vision Transformer originally pre-trained on \texttt{ImageNet-21K} for natural image recognition tasks \citep{liu2021swin, carion2020end, wang2021max, chen2021transformer}. Basically, the model first partitions each input image into fixed-size patches, which are embedded and processed through a series of self-attention blocks with a hierarchical structure. This enables the model to capture both local patterns and global dependencies across the light curve image. After processing the image through the \textsc{SwinV2} backbone, an adaptive average pooling layer computes the mean across all tokens, producing a single pooled embedding that is used for classification. Further architectural details and an illustration of the model are provided in Appendix~\ref{appendix:swinv2}. To assess the effectiveness of this vision-based approach, we compare it with the Astronomical Transformer for time series And Tabular data \citep[ATAT;][]{cabrera2024atat}, a specialized Transformer architecture tailored for sequential modeling of multi-band astronomical time series.

\subsection{Multi-Survey Methodology}
\label{sec:ms_methodology}

\begin{figure}[t]
\vskip 0.2in
\begin{center}
\centerline{\includegraphics[width=8cm]{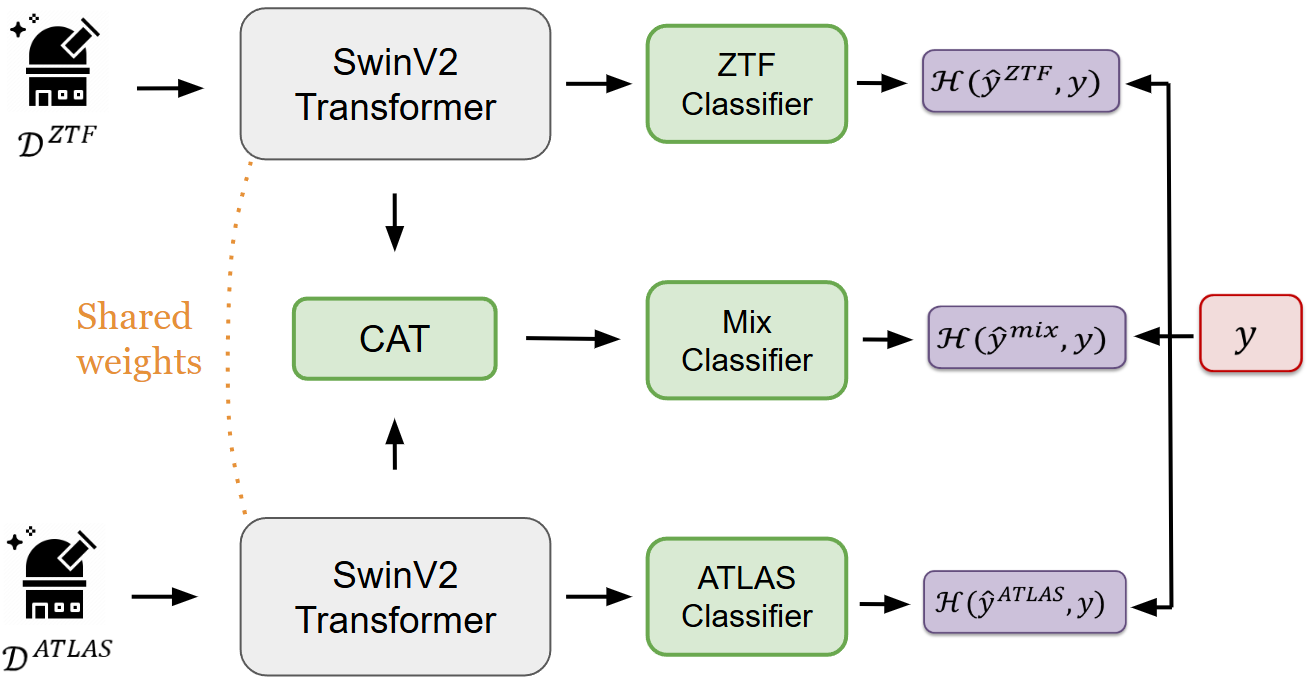}}
\caption{Multi-survey architecture.}
\label{fig:ms_arch}
\end{center}
\vskip -0.4in
\end{figure}

We explore multiple strategies for combining ZTF and ATLAS data to improve classification performance. As a baseline, we first train separate single-survey \textsc{SwinV2} models for ZTF and ATLAS. We then perform prediction-level fusion by averaging the predicted probabilities of these two models to obtain a combined prediction. Additionally, we perform embedding-level fusion by extracting embeddings from the single-survey models and training a Balanced Random Forest (BRF) classifier on the concatenated embeddings. We also experiment with input-level fusion by creating a new dataset in which light curves from both surveys are represented as multi-band images in which bands $g$, $r$, $o$, and $c$ are concatenated spatially into a single image, as shown in Figure~\ref{fig:input_model} (stacked input). Finally, we propose a multi-survey architecture that simultaneously processes photometric data from both surveys. Figure~\ref{fig:ms_arch} illustrates the overall architecture of this approach. Specifically, light curves from ZTF (\( \mathcal{D}^{\mathrm{ZTF}} \)) and ATLAS (\( \mathcal{D}^{\mathrm{ATLAS}} \)) are treated as independent input streams and processed by two parallel \textsc{SwinV2} Transformer branches with shared weights. The output of each branch is connected to a dedicated linear classifier, ZTF Classifier and ATLAS Classifier, that produces survey-specific predictions \( \hat{y}^{\mathrm{ZTF}} \) and \( \hat{y}^{\mathrm{ATLAS}} \), each optimized using a cross-entropy loss \( \mathcal{H}(\hat{y}, y) \). In parallel, the pooled embeddings from both streams are concatenated (CAT) and passed through a Mix Classifier, and we experiment with both a Linear Layer and a Multi-Layer Perceptron (MLP) implementation for this classifier, to generate a joint prediction \( \hat{y}^{\mathrm{mix}} \), also optimized via cross-entropy. The total loss is computed as $\mathcal{L}_{\text{total}} = \mathcal{H}(\hat{y}^{\mathrm{ZTF}}, y) + \mathcal{H}(\hat{y}^{\mathrm{ATLAS}}, y) + \mathcal{H}(\hat{y}^{\mathrm{mix}}, y)$. This design enables the model to jointly learn survey-specific patterns and cross-survey interactions, while maintaining flexibility to output individual and joint predictions.

\section{Experiments}
\label{sec:experiments}

\subsection{Data Description}
\label{sec:data_description}
We use forced-photometry light curves from ZTF and ATLAS. Forced photometry retrieves fluxes at fixed sky positions, enabling the detection of faint signals. ZTF objects were selected from the public alert stream and processed using the Automatic Learning for the Rapid Classification of Events system \citep[ALeRCE;][]{forster2021automatic}. Historical fluxes were obtained via the ZTF Forced Photometry Service \citep{masci2023new}, using each object's RA and Dec, with standard quality filtering. Forced photometry complements the alert stream with more complete flux information. This type of data has been included in ZTF alerts only since October 2023. Labels were assigned through crossmatch with external catalogs and expert validation. Preprocessing removed epochs with negative flux uncertainties or bad-quality flags, selected forced photometry observations from up to 30 days before the first alert through to the last alert, discarded objects with fewer than eight detections, and filtered spurious events. We used calibrated differential fluxes in microjanskys ($\mu$Jy). A forthcoming ALeRCE publication will describe the dataset in detail. After preprocessing, the ZTF dataset included 39,483 light curves in $g$ and $r$ bands spanning 21 classes, with a median cadence of 1.13 days and a standard deviation of 6.5 days, considering both bands. Data were split into training, validation, and test sets with class-stratified sampling: $20\%$ for testing, and the remaining $80\%$ split into five training folds.

For the multi-survey approach, we incorporated ATLAS forced-photometry data. Using the RA and Dec of each ZTF object, we queried the ATLAS service \citep{tonry2018atlas, heinze2018first, smith2020design, shingles2021release} to retrieve historical fluxes at the same positions. ATLAS performs PSF-fit forced photometry on difference images using \texttt{tphot} \citep{tonry2010early}. To ensure quality, we retained only measurements with $0.5 < \chi^{2}/N < 3$ and sky magnitudes above 18 in the $o$ band and 18.5 in the $c$ band, following the criteria from \citet{silva2024physical}. Fluxes are reported in $\mu$Jy in the AB system, with the filter ($c$ or $o$) indicated for each observation. The ATLAS dataset comprised 39,464 light curves, with the same classes and data split as the ZTF dataset. It has a median cadence of 1.52 days with a standard deviation of 11.03 days, considering both bands. The number of objects per class and the class taxonomy are shown in Table~\ref{table:number_of_objects_groups}. The table also presents the division into transient, stochastic, and periodic sources.

\begin{table}[t]
\caption{Number of objects per class in ZTF, grouped as \textit{Transient}, \textit{Stochastic}, and \textit{Periodic}, as defined by the ALeRCE Team.}
\label{table:number_of_objects_groups}
\vskip 0.1in
\begin{center}
\renewcommand{\arraystretch}{1.2}
\begin{footnotesize}
\begin{tabular}{|l|l|l|}
\toprule
\textbf{Transient} & \textbf{Stochastic} & \textbf{Periodic} \\
\midrule
SNIa (2486)       & Microlensing (29)    & LPV (2931)            \\
SESN (565)        & QSO (2923)           & EA (2909)             \\
SNII (1393)       & AGN (2898)           & EB/EW (2921)        \\
SNIIn (267)       & Blazar (1393)        & Periodic-Other (1269) \\
SLSN (185)        & YSO (2773)           & RSCVn (2452)          \\
TDE (81)          & CV/Nova (1640)       & CEP (2891)            \\
-                 &                      & RRLab (2931)          \\
-                 &                      & RRLc (2936)           \\
-                 &                      & DSCT (1610)           \\
\bottomrule
\end{tabular}
\end{footnotesize}
\end{center}
\vskip -0.3in
\end{table}

\subsection{Training Details}
\label{sec:training_details}

In our experiments, light curves were converted to images with \texttt{Matplotlib}, transformed into RGB tensors, and adjusted to minimize whitespace. We used the \textsc{SwinV2} model pre-trained on \texttt{ImageNet-21K}, available via \texttt{Hugging Face}\footnote{\href{https://huggingface.co/microsoft/swinv2-tiny-patch4-window16-256}{Pre-trained \texttt{Hugging Face} model used.}}. The model operates on $256 {\times} 256 {\times} 3$ images, divides them into $4 {\times} 4$ patches, uses an embedding dimension of $C=96$, and performs self-attention within $16 {\times} 16$ windows. Image processing used \texttt{AutoImageProcessor}, including resizing to $256 {\times} 256$, rescaling to $[0,1]$, and normalization with \texttt{ImageNet-21K} statistics. Training was conducted on an NVIDIA RTX A5000 GPU.

We used a marker size of 1.0, a line width of 2.0, flux errors, and a learning rate of $5 \cdot 10^{-6}$ as hyperparameters for image generation, following \citet{moreno2025leveraging}, who found this configuration to be optimal for the two bands in the public Catalog of Variable Stars from the Massive Compact Halo Object (MACHO) survey. To mitigate class imbalance, we employed a weighted sampling strategy\footnote{\protect\href{https://pytorch.org/docs/stable/data.html\#torch.utils.data.WeightedRandomSampler}{PyTorch's WeightedRandomSampler.}} based on class frequencies, promoting a more balanced representation and better generalization. The model was optimized with Adam and early stopping (patience: 5 epochs) using validation F1-score.

\section{Results}
\label{sec:results}

Table~\ref{table:fusion_strategies} summarizes the classification performance on the test set, in terms of F1-score, for the different strategies explored for combining ZTF and ATLAS data. As single-survey baselines, we compare the ATAT and \textsc{SwinV2} models trained independently on ZTF and ATLAS. ATAT is a specialized Transformer architecture designed to process multi-band light curves jointly with tabular data. In this work, however, we focus exclusively on light curves and therefore do not use metadata or derived features from the light curves, to ensure a fair comparison with the vision Transformer. This configuration is referred to as ATAT (LC). We prepared the input data for ATAT using sliding windows of 200 observations per light curve and per band, as the quadratic complexity of the attention mechanism limits the feasible input length. The median number of observations per light curve is 952.0 for ZTF and 1493.0 for ATLAS. Final class probabilities were obtained by averaging the predicted probability vectors across all windows for each light curve. Table~\ref{table:fusion_strategies} shows that ATAT achieved F1-scores of 57.0\% (ZTF, Model~1) and 49.6\% (ATLAS, Model~2), while \textsc{SwinV2} reached higher F1-scores of 64.9\% (ZTF, Model~3) and 54.0\% (ATLAS, Model~4). These results indicate that distinguishing classes on ZTF is easier for both models than on ATLAS, likely due to differences in filter sets, cadence, and photometric noise. They also highlight the potential of vision Transformers for multi-band light curve classification compared to the specialized ATAT architecture. We adopt \textsc{SwinV2} as the baseline for subsequent fusion experiments, as it achieved the best overall performance.

\begin{table}[t]
\caption{Comparison of ATAT and \textsc{SwinV2}-based strategies for photometric classification using ZTF and ATLAS data.}
\label{table:fusion_strategies}
\vskip 0.15in
\begin{center}
\begin{small}
\begin{sc}
\begin{tabular}{c l c}
\toprule
Model & Description & F1-score \\
\midrule
1 & ATAT (LC) (ZTF only) & 57.0 $\pm$ 1.4 \\ 
2 & ATAT (LC) (ATLAS only) & 49.6 $\pm$ 1.9 \\ 
3 & \textsc{SwinV2} (ZTF only) & 64.9 $\pm$ 1.6 \\ 
4 & \textsc{SwinV2} (ATLAS only) & 54.0 $\pm$ 1.3 \\ \midrule 
3 + 4 & Averaged probabilities & 68.9 $\pm$ 1.3 \\ 
3 + 4 & BRF on embeddings & 68.7 $\pm$ 0.9 \\ \midrule 
5 & \textsc{SwinV2} (Stacked input) & 59.0 $\pm$ 1.3 \\ 
6 & MS-\textsc{SwinV2}-Linear & \textbf{69.9} $\pm$ \textbf{1.7} \\
7 & MS-\textsc{SwinV2}-MLP & 69.8 $\pm$ 1.5 \\
\bottomrule
\end{tabular}
\end{sc}
\end{small}
\end{center}
\vskip -0.25in
\end{table}

When combining the two single-survey models, both prediction-level fusion (averaged probabilities) and embedding-level fusion (BRF on embeddings) achieved higher F1-scores than the individual baselines, reaching 68.9\% and 68.7\%, respectively. A hyperparameter search is reported in Table~\ref{table:rf_hyperparam_tuning} of Appendix~\ref{appendix:hyperparameters} for the BRF classifier. These results indicate that even simple post-hoc fusion strategies can leverage complementary information between surveys. For input-level fusion, where light curves from both surveys are represented as stacked multi-band images (Figure~\ref{fig:input_model}, stacked input) and processed by \textsc{SwinV2} (Model~5), reached only 59.0\%, below the single-survey ZTF baseline. This suggests that naive early fusion may introduce additional complexity or noise when combining heterogeneous light curves into a single input.

Finally, the proposed multi-survey \textsc{SwinV2} architectures, which process each survey as an independent input stream with shared weights, achieved the highest performance among the strategies evaluated. The Linear variant (Model~6) reached 69.9\%, and the MLP variant (Model~7) 69.8\% F1-score. Although ATLAS-only models perform worse than ZTF-only models, incorporating ATLAS data in the multi-survey framework provides additional signals that improve overall classification. In contrast to post-hoc fusion (e.g., 68.9\% with averaged probabilities), the multi-survey approach captures cross-survey interactions during training. This is reflected in the confusion matrices shown in Figure~\ref{fig:cm_models} (Appendix~\ref{appendix:cm_matrices}). Moreover, it is well-suited for scaling to additional surveys, simplifies deployment with a unified model, and offers a flexible foundation for incorporating other modalities, such as metadata, contextual features, or spectra.

These results suggest the benefit of modeling survey-specific representations while jointly capturing cross-survey interactions within a unified architecture. Using a pre-trained vision Transformer also avoids the complexity of designing specialized representations for multi-band light curve data, offering a simple alternative to classical sequence models.

\section{Conclusion}
\label{sec:conclusion}

In this work, we presented a systematic exploration of fusion strategies for photometric classification using multi-survey light curves from ZTF and ATLAS. Our results indicate that multi-survey architectures, which process each survey as an independent input, obtained the highest classification performance among the approaches evaluated, compared to single-survey baselines and simpler fusion methods. These observations suggest that incorporating information from additional telescopes can be beneficial for classification, even when the data are noisier or of lower individual quality. The results also point to the importance of explicitly modeling survey-specific characteristics when integrating heterogeneous time-domain data. As multi-survey and multi-instrument data become increasingly common, such architectures may provide a useful foundation for scalable and adaptable photometric classifiers. Future work will extend this approach to incorporate metadata and additional features \citep{cabrera2024atat}, and to explore its use in real-time alert processing.

\section*{Acknowledgements}

The authors acknowledge support from the National Agency for Research and Development (ANID) grants: FONDECYT Regular 1231877 (DMC, GCV); FONDECYT Regular 1241005 (FEB); FONDECYT Regular 1231637 (MC); FONDECYT Iniciación 11241477 (LHG); FONDECYT 1220829 (PAE); Millennium Science Initiative Program ICN12\_009 (GCV, AMMA); Millennium Science Initiative Program AIM23\_0001, awarded to the Millennium Institute of Astrophysics (FEB, FF, LHG, MC, PAE, AMMA); Millennium Science Initiative Program NCN2023\_002 (LHG). This work was also supported by the BASAL Center of Mathematical Modeling Grant FB210005 (AMMA, FF), ANID-Chile BASAL CATA FB210003 (FEB), ANID's BASAL project FB210003 (MC), and the Quimal infrastructure funds QUIMAL190012 and QUIMAL240008 (FF).

\bibliography{example_paper}

\begin{thebibliography}{50}
\providecommand{\natexlab}[1]{#1}
\providecommand{\url}[1]{\texttt{#1}}
\expandafter\ifx\csname urlstyle\endcsname\relax
  \providecommand{\doi}[1]{doi: #1}\else
  \providecommand{\doi}{doi: \begingroup \urlstyle{rm}\Url}\fi

\bibitem[Aguirre et~al.(2019)Aguirre, Pichara, and Becker]{aguirre2019deep}
Aguirre, C., Pichara, K., and Becker, I.
\newblock Deep multi-survey classification of variable stars.
\newblock \emph{Monthly Notices of the Royal Astronomical Society}, 482\penalty0 (4):\penalty0 5078--5092, 2019.

\bibitem[Allam~Jr \& McEwen(2024)Allam~Jr and McEwen]{allam2024paying}
Allam~Jr, T. and McEwen, J.~D.
\newblock Paying attention to astronomical transients: introducing the time-series transformer for photometric classification.
\newblock \emph{RAS Techniques and Instruments}, 3\penalty0 (1):\penalty0 209--223, 2024.

\bibitem[Becker et~al.(2025)Becker, Protopapas, Catelan, and Pichara]{becker2025}
Becker, I., Protopapas, P., Catelan, M., and Pichara, K.
\newblock Multiband embeddings of light curves.
\newblock \emph{A\&A}, 694:\penalty0 A183, 2025.
\newblock \doi{10.1051/0004-6361/202347461}.
\newblock URL \url{https://doi.org/10.1051/0004-6361/202347461}.

\bibitem[Bellm et~al.(2018)Bellm, Kulkarni, Graham, Dekany, Smith, Riddle, Masci, Helou, Prince, Adams, et~al.]{bellm2018zwicky}
Bellm, E.~C., Kulkarni, S.~R., Graham, M.~J., Dekany, R., Smith, R.~M., Riddle, R., Masci, F.~J., Helou, G., Prince, T.~A., Adams, S.~M., et~al.
\newblock The zwicky transient facility: system overview, performance, and first results.
\newblock \emph{Publications of the Astronomical Society of the Pacific}, 131\penalty0 (995):\penalty0 018002, 2018.

\bibitem[Boone(2019)]{boone2019avocado}
Boone, K.
\newblock Avocado: Photometric classification of astronomical transients with gaussian process augmentation.
\newblock \emph{The Astronomical Journal}, 158\penalty0 (6):\penalty0 257, 2019.

\bibitem[Cabrera-Vives et~al.(2024)Cabrera-Vives, Moreno-Cartagena, Astorga, Reyes-Jainaga, F{\"o}rster, Huijse, Arredondo, Arancibia, Bayo, Catelan, et~al.]{cabrera2024atat}
Cabrera-Vives, G., Moreno-Cartagena, D., Astorga, N., Reyes-Jainaga, I., F{\"o}rster, F., Huijse, P., Arredondo, J., Arancibia, A.~M., Bayo, A., Catelan, M., et~al.
\newblock Atat: Astronomical transformer for time series and tabular data.
\newblock \emph{Astronomy \& Astrophysics}, 689:\penalty0 A289, 2024.

\bibitem[C{\'a}diz-Leyton et~al.(2024{\natexlab{a}})C{\'a}diz-Leyton, Cabrera-Vives, Protopapas, Moreno-Cartagena, and Donoso-Oliva]{cadiz2024workshop}
C{\'a}diz-Leyton, M., Cabrera-Vives, G., Protopapas, P., Moreno-Cartagena, D., and Donoso-Oliva, C.
\newblock Transformer-based astronomical time series model with uncertainty estimation for detecting misclassified instances.
\newblock In \emph{LatinX in AI workshop, 41 st International Conference on Machine Learning ({ICML}), PMLR 235}, Vienna, Austria, 2024{\natexlab{a}}.
\newblock URL \url{https://arxiv.org/abs/2411.01363}.

\bibitem[C{\'a}diz-Leyton et~al.(2024{\natexlab{b}})C{\'a}diz-Leyton, Cabrera-Vives, Protopapas, Moreno-Cartagena, Donoso-Oliva, and Becker]{cadiz2024Journal}
C{\'a}diz-Leyton, M., Cabrera-Vives, G., Protopapas, P., Moreno-Cartagena, D., Donoso-Oliva, C., and Becker, I.
\newblock Uncertainty estimation for time series classification: Exploring predictive uncertainty in transformer-based models for variable stars.
\newblock \emph{Astronomy \& Astrophysics}, 2024{\natexlab{b}}.
\newblock Accepted for publication.

\bibitem[Carion et~al.(2020)Carion, Massa, Synnaeve, Usunier, Kirillov, and Zagoruyko]{carion2020end}
Carion, N., Massa, F., Synnaeve, G., Usunier, N., Kirillov, A., and Zagoruyko, S.
\newblock End-to-end object detection with transformers.
\newblock In \emph{European conference on computer vision}, pp.\  213--229. Springer, 2020.

\bibitem[Carrasco-Davis et~al.(2019)Carrasco-Davis, Cabrera-Vives, F{\"o}rster, Est{\'e}vez, Huijse, Protopapas, Reyes, Mart{\'\i}nez-Palomera, and Donoso]{carrasco2019deep}
Carrasco-Davis, R., Cabrera-Vives, G., F{\"o}rster, F., Est{\'e}vez, P.~A., Huijse, P., Protopapas, P., Reyes, I., Mart{\'\i}nez-Palomera, J., and Donoso, C.
\newblock Deep learning for image sequence classification of astronomical events.
\newblock \emph{Publications of the Astronomical Society of the Pacific}, 131\penalty0 (1004):\penalty0 108006, 2019.

\bibitem[{Catelan}(2023)]{Catelan}
{Catelan}, M.
\newblock {Stellar Variability in Ground-Based Photometric Surveys: AN Overview}.
\newblock In \emph{Memorie della Societa Astronomica Italiana}, volume~94, pp.\ ~56, December 2023.
\newblock \doi{10.36116/MEMSAIT_94N4.2023.56}.

\bibitem[Chen et~al.(2021)Chen, Yan, Zhu, Wang, Yang, and Lu]{chen2021transformer}
Chen, X., Yan, B., Zhu, J., Wang, D., Yang, X., and Lu, H.
\newblock Transformer tracking.
\newblock In \emph{Proceedings of the IEEE/CVF conference on computer vision and pattern recognition}, pp.\  8126--8135, 2021.

\bibitem[Donoso-Oliva et~al.(2023)Donoso-Oliva, Becker, Protopapas, Cabrera-Vives, Vishnu, and Vardhan]{donoso2023astromer}
Donoso-Oliva, C., Becker, I., Protopapas, P., Cabrera-Vives, G., Vishnu, M., and Vardhan, H.
\newblock Astromer-a transformer-based embedding for the representation of light curves.
\newblock \emph{Astronomy \& Astrophysics}, 670:\penalty0 A54, 2023.

\bibitem[Drake et~al.(2012)Drake, Djorgovski, Mahabal, Prieto, Beshore, Graham, Catalan, Larson, Christensen, and Donalek]{drake2012catalina}
Drake, A.~J., Djorgovski, S.~G., Mahabal, A., Prieto, J.~L., Beshore, E., Graham, M.~J., Catalan, M., Larson, S., Christensen, E., and Donalek, C.
\newblock The catalina real-time transient survey.
\newblock In \emph{Proceedings of the International Astronomical Union, Symposium S285: New Horizons in Time-Domain Astronomy}, volume~7, pp.\  306--308. Cambridge University Press, 2012.
\newblock \doi{10.1017/S1743921312000889}.

\bibitem[Engel et~al.(2024)Engel, Narayan, and Byler]{engel2024preliminary}
Engel, A., Narayan, G., and Byler, N.
\newblock Preliminary report on mantis shrimp: a multi‑survey computer vision photometric redshift model.
\newblock \emph{arXiv preprint arXiv:2402.03535}, 2024.
\newblock URL \url{https://arxiv.org/abs/2402.03535}.
\newblock Submitted to AI4Differential Equations in Science Workshop, ICLR 2024.

\bibitem[F{\"o}rster et~al.(2021)F{\"o}rster, Cabrera-Vives, Castillo-Navarrete, Est{\'e}vez, S{\'a}nchez-S{\'a}ez, Arredondo, Bauer, Carrasco-Davis, Catelan, Elorrieta, et~al.]{forster2021automatic}
F{\"o}rster, F., Cabrera-Vives, G., Castillo-Navarrete, E., Est{\'e}vez, P., S{\'a}nchez-S{\'a}ez, P., Arredondo, J., Bauer, F., Carrasco-Davis, R., Catelan, M., Elorrieta, F., et~al.
\newblock The automatic learning for the rapid classification of events (alerce) alert broker.
\newblock \emph{The Astronomical Journal}, 161\penalty0 (5):\penalty0 242, 2021.

\bibitem[{Gaia Collaboration} et~al.(2023){Gaia Collaboration}, Vallenari, Brown, Prusti, de~Bruijne, and et~al.]{gaia2023gaia}
{Gaia Collaboration}, Vallenari, A., Brown, A. G.~A., Prusti, T., de~Bruijne, J. H.~J., and et~al.
\newblock Gaia data release 3: Summary of the content and survey properties.
\newblock \emph{Astronomy \& Astrophysics}, 674:\penalty0 A1, 2023.
\newblock \doi{10.1051/0004-6361/202243940}.

\bibitem[Garc{\'\i}a-Jara et~al.(2022)Garc{\'\i}a-Jara, Protopapas, and Est{\'e}vez]{garcia2022improving}
Garc{\'\i}a-Jara, G., Protopapas, P., and Est{\'e}vez, P.~A.
\newblock Improving astronomical time-series classification via data augmentation with generative adversarial networks.
\newblock \emph{The Astrophysical Journal}, 935\penalty0 (1):\penalty0 23, 2022.

\bibitem[G{\'o}mez et~al.(2020)G{\'o}mez, Neira, Hern{\'a}ndez~Hoyos, Arbel{\'a}ez, and Forero-Romero]{gomez2020classifying}
G{\'o}mez, C., Neira, M., Hern{\'a}ndez~Hoyos, M., Arbel{\'a}ez, P., and Forero-Romero, J.~E.
\newblock Classifying image sequences of astronomical transients with deep neural networks.
\newblock \emph{Monthly Notices of the Royal Astronomical Society}, 499\penalty0 (3):\penalty0 3130--3138, 2020.

\bibitem[Gupta \& Muthukrishna(2025)Gupta and Muthukrishna]{gupta2025transfer}
Gupta, R. and Muthukrishna, D.
\newblock Transfer learning for transient classification: From simulations to real data and ztf to lsst.
\newblock \emph{arXiv preprint arXiv:2502.18558}, 2025.
\newblock Submitted to MNRAS.

\bibitem[Heinze et~al.(2018)Heinze, Tonry, Denneau, Flewelling, Stalder, Rest, Smith, Smartt, and Weiland]{heinze2018first}
Heinze, A., Tonry, J.~L., Denneau, L., Flewelling, H., Stalder, B., Rest, A., Smith, K.~W., Smartt, S.~J., and Weiland, H.
\newblock A first catalog of variable stars measured by the asteroid terrestrial-impact last alert system (atlas).
\newblock \emph{The Astronomical Journal}, 156\penalty0 (5):\penalty0 241, 2018.

\bibitem[Jamal \& Bloom(2020)Jamal and Bloom]{jamal2020neural}
Jamal, S. and Bloom, J.~S.
\newblock On neural architectures for astronomical time-series classification with application to variable stars.
\newblock \emph{The Astrophysical Journal Supplement Series}, 250\penalty0 (2):\penalty0 30, 2020.

\bibitem[Kaiser et~al.(2010)Kaiser, Burgett, Chambers, Denneau, Heasley, Jedicke, Magnier, Morgan, Onaka, and Tonry]{kaiser2010pan}
Kaiser, N., Burgett, W., Chambers, K., Denneau, L., Heasley, J., Jedicke, R., Magnier, E., Morgan, J., Onaka, P., and Tonry, J.
\newblock The pan-starrs wide-field optical/nir imaging survey.
\newblock In \emph{Ground-based and airborne telescopes III}, volume 7733, pp.\  159--172. SPIE, 2010.

\bibitem[Liu et~al.(2021)Liu, Lin, Cao, Hu, Wei, Zhang, Lin, and Guo]{liu2021swin}
Liu, Z., Lin, Y., Cao, Y., Hu, H., Wei, Y., Zhang, Z., Lin, S., and Guo, B.
\newblock Swin transformer: Hierarchical vision transformer using shifted windows.
\newblock In \emph{Proceedings of the IEEE/CVF international conference on computer vision}, pp.\  10012--10022, 2021.

\bibitem[Liu et~al.(2022)Liu, Hu, Lin, Yao, Xie, Wei, Ning, Cao, Zhang, Dong, et~al.]{liu2022swin}
Liu, Z., Hu, H., Lin, Y., Yao, Z., Xie, Z., Wei, Y., Ning, J., Cao, Y., Zhang, Z., Dong, L., et~al.
\newblock Swin transformer v2: Scaling up capacity and resolution.
\newblock In \emph{Proceedings of the IEEE/CVF conference on computer vision and pattern recognition}, pp.\  12009--12019, 2022.

\bibitem[Lochner et~al.(2016)Lochner, McEwen, Peiris, Lahav, and Winter]{lochner2016photometric}
Lochner, M., McEwen, J.~D., Peiris, H.~V., Lahav, O., and Winter, M.~K.
\newblock Photometric supernova classification with machine learning.
\newblock \emph{The Astrophysical Journal Supplement Series}, 225\penalty0 (2):\penalty0 31, 2016.

\bibitem[Mahabal et~al.(2017)Mahabal, Sheth, Gieseke, Pai, Djorgovski, Drake, and Graham]{mahabal2017deep}
Mahabal, A., Sheth, K., Gieseke, F., Pai, A., Djorgovski, S.~G., Drake, A.~J., and Graham, M.~J.
\newblock Deep-learnt classification of light curves.
\newblock In \emph{2017 IEEE symposium series on computational intelligence (SSCI)}, pp.\  1--8. IEEE, 2017.

\bibitem[Masci et~al.(2023)Masci, Laher, Rusholme, Shupe, Paladini, Groom, Wold, Miller, and Drake]{masci2023new}
Masci, F.~J., Laher, R.~R., Rusholme, B., Shupe, D., Paladini, R., Groom, S., Wold, A., Miller, A.~A., and Drake, A.
\newblock A new forced photometry service for the zwicky transient facility.
\newblock \emph{arXiv preprint arXiv:2305.16279}, 2023.

\bibitem[Monsalves et~al.(2024)Monsalves, Jaque~Arancibia, Bayo, S{\'a}nchez-S{\'a}ez, Angeloni, Damke, and Segura Van~de Perre]{monsalves2024application}
Monsalves, N., Jaque~Arancibia, M., Bayo, A., S{\'a}nchez-S{\'a}ez, P., Angeloni, R., Damke, G., and Segura Van~de Perre, J.
\newblock Application of convolutional neural networks to time domain astrophysics. 2d image analysis of ogle light curves.
\newblock \emph{Astronomy \& Astrophysics}, 691:\penalty0 A106, 2024.

\bibitem[Moreno-Cartagena et~al.(2023)Moreno-Cartagena, Cabrera-Vives, Protopapas, Donoso-Oliva, P{\'e}rez-Carrasco, and C{\'a}diz-Leyton]{moreno2023positional}
Moreno-Cartagena, D., Cabrera-Vives, G., Protopapas, P., Donoso-Oliva, C., P{\'e}rez-Carrasco, M., and C{\'a}diz-Leyton, M.
\newblock Positional encodings for light curve transformers: Playing with positions and attention.
\newblock In \emph{Machine Learning for Astrophysics Workshop, 40th International Conference on Machine Learning ({ICML}), PMLR 202}, Honolulu, Hawaii, USA, 2023.
\newblock URL \url{https://ml4astro.github.io/icml2023/assets/55.pdf}.

\bibitem[Moreno-Cartagena et~al.(2025)Moreno-Cartagena, Protopapas, Cabrera-Vives, C{\'a}diz-Leyton, Becker, and Donoso-Oliva]{moreno2025leveraging}
Moreno-Cartagena, D., Protopapas, P., Cabrera-Vives, G., C{\'a}diz-Leyton, M., Becker, I., and Donoso-Oliva, C.
\newblock Leveraging pre-trained visual transformers for multi-band photometric light curve classification.
\newblock \emph{arXiv preprint arXiv:2502.20479}, 2025.
\newblock Submitted to Astronomy \& Astrophysics.

\bibitem[Murray et~al.()Murray, Fabbro, Herzberg, and Yi]{murrayunsupervised}
Murray, C., Fabbro, S., Herzberg, N., and Yi, K.~M.
\newblock Unsupervised learning of deep features through best-fits for observational cosmology.

\bibitem[Muthukrishna et~al.(2019)Muthukrishna, Narayan, Mandel, Biswas, and Hlo{\v{z}}ek]{muthukrishna2019rapid}
Muthukrishna, D., Narayan, G., Mandel, K.~S., Biswas, R., and Hlo{\v{z}}ek, R.
\newblock Rapid: early classification of explosive transients using deep learning.
\newblock \emph{Publications of the Astronomical Society of the Pacific}, 131\penalty0 (1005):\penalty0 118002, 2019.

\bibitem[Pan et~al.(2022)Pan, Ting, and Yu]{pan2022astroconformer}
Pan, J., Ting, Y.-S., and Yu, J.
\newblock Astroconformer: Inferring surface gravity of stars from stellar light curves with transformer.
\newblock In \emph{Machine Learning for Astrophysics Workshop, 39th International Conference on Machine Learning ({ICML}), PMLR 162}, Baltimore, Maryland, USA, 2022.

\bibitem[Pimentel et~al.(2022)Pimentel, Est{\'e}vez, and F{\"o}rster]{pimentel2022deep}
Pimentel, {\'O}., Est{\'e}vez, P.~A., and F{\"o}rster, F.
\newblock Deep attention-based supernovae classification of multiband light curves.
\newblock \emph{The Astronomical Journal}, 165\penalty0 (1):\penalty0 18, 2022.

\bibitem[Rizhko \& Bloom(2024{\natexlab{a}})Rizhko and Bloom]{rizhko2024astrom}
Rizhko, M. and Bloom, J.~S.
\newblock Self-supervised multimodal model for astronomy.
\newblock In \emph{Neurips 2024 Workshop Foundation Models for Science: Progress, Opportunities, and Challenges}, 2024{\natexlab{a}}.

\bibitem[Rizhko \& Bloom(2024{\natexlab{b}})Rizhko and Bloom]{rizhko2024self}
Rizhko, M. and Bloom, J.~S.
\newblock Astrom$\textsuperscript{3}$: A self-supervised multimodal model for astronomy.
\newblock \emph{arXiv preprint arXiv:2411.08842}, 2024{\natexlab{b}}.
\newblock Submitted to MNRAS.

\bibitem[S{\'a}nchez-S{\'a}ez et~al.(2021)S{\'a}nchez-S{\'a}ez, Reyes, Valenzuela, F{\"o}rster, Eyheramendy, Elorrieta, Bauer, Cabrera-Vives, Est{\'e}vez, Catelan, et~al.]{sanchez2021alert}
S{\'a}nchez-S{\'a}ez, P., Reyes, I., Valenzuela, C., F{\"o}rster, F., Eyheramendy, S., Elorrieta, F., Bauer, F., Cabrera-Vives, G., Est{\'e}vez, P., Catelan, M., et~al.
\newblock Alert classification for the alerce broker system: The light curve classifier.
\newblock \emph{The Astronomical Journal}, 161\penalty0 (3):\penalty0 141, 2021.

\bibitem[Shappee et~al.(2014)Shappee, Prieto, Grupe, Kochanek, Stanek, De~Rosa, Mathur, Zu, Peterson, Pogge, et~al.]{shappee2014man}
Shappee, B.~J., Prieto, J., Grupe, D., Kochanek, C., Stanek, K., De~Rosa, G., Mathur, S., Zu, Y., Peterson, B., Pogge, R., et~al.
\newblock The man behind the curtain: X-rays drive the uv through nir variability in the 2013 active galactic nucleus outburst in ngc 2617.
\newblock \emph{The Astrophysical Journal}, 788\penalty0 (1):\penalty0 48, 2014.

\bibitem[Shingles et~al.(2021)Shingles, Smith, Young, Smartt, Tonry, Denneau, Heinze, Weiland, Flewelling, Stalder, et~al.]{shingles2021release}
Shingles, L., Smith, K., Young, D., Smartt, S., Tonry, J., Denneau, L., Heinze, A., Weiland, H., Flewelling, H., Stalder, B., et~al.
\newblock Release of the atlas forced photometry server for public use.
\newblock \emph{Transient Name Server AstroNote 2021-7}, 7:\penalty0 1--7, 2021.

\bibitem[Silva-Farf{\'a}n et~al.(2024)Silva-Farf{\'a}n, F{\"o}rster, Moriya, Hern{\'a}ndez-Garc{\'\i}a, Arancibia, S{\'a}nchez-S{\'a}ez, Anderson, Tonry, and Clocchiatti]{silva2024physical}
Silva-Farf{\'a}n, J., F{\"o}rster, F., Moriya, T.~J., Hern{\'a}ndez-Garc{\'\i}a, L., Arancibia, A.~M., S{\'a}nchez-S{\'a}ez, P., Anderson, J.~P., Tonry, J.~L., and Clocchiatti, A.
\newblock Physical properties of type ii supernovae inferred from ztf and atlas photometric data.
\newblock \emph{The Astrophysical Journal}, 969\penalty0 (1):\penalty0 57, 2024.

\bibitem[Smith et~al.(2020)Smith, Smartt, Young, Tonry, Denneau, Flewelling, Heinze, Weiland, Stalder, Rest, et~al.]{smith2020design}
Smith, K., Smartt, S., Young, D., Tonry, J., Denneau, L., Flewelling, H., Heinze, A., Weiland, H., Stalder, B., Rest, A., et~al.
\newblock Design and operation of the atlas transient science server.
\newblock \emph{Publications of the Astronomical Society of the Pacific}, 132\penalty0 (1014):\penalty0 085002, 2020.

\bibitem[{Starr} et~al.(2012){Starr}, {Richards}, {Brink}, {Miller}, {Bloom}, {Butler}, {James}, and {Long}]{RichardsMS}
{Starr}, D.~L., {Richards}, J.~W., {Brink}, H., {Miller}, A.~A., {Bloom}, J.~S., {Butler}, N.~R., {James}, J.~B., and {Long}, J.~P.
\newblock {ALLStars: Overcoming Multi-Survey Selection Bias using Crowd-Sourced Active Learning}.
\newblock In {Ballester}, P., {Egret}, D., and {Lorente}, N.~P.~F. (eds.), \emph{Astronomical Data Analysis Software and Systems XXI}, volume 461 of \emph{Astronomical Society of the Pacific Conference Series}, pp.\  581, September 2012.

\bibitem[Tonry et~al.(2018)Tonry, Denneau, Heinze, Stalder, Smith, Smartt, Stubbs, Weiland, and Rest]{tonry2018atlas}
Tonry, J., Denneau, L., Heinze, A., Stalder, B., Smith, K., Smartt, S., Stubbs, C., Weiland, H., and Rest, A.
\newblock Atlas: a high-cadence all-sky survey system.
\newblock \emph{Publications of the Astronomical Society of the Pacific}, 130\penalty0 (988):\penalty0 064505, 2018.

\bibitem[Tonry(2010)]{tonry2010early}
Tonry, J.~L.
\newblock An early warning system for asteroid impact.
\newblock \emph{Publications of the Astronomical Society of the Pacific}, 123\penalty0 (899):\penalty0 58, 2010.

\bibitem[Udalski et~al.(2015)Udalski, Szymański, and Szymański]{Udalski2015}
Udalski, A., Szymański, M.~K., and Szymański, G.
\newblock Ogle-iv: Fourth phase of the optical gravitational lensing experiment.
\newblock \emph{Acta Astronomica}, 65\penalty0 (1):\penalty0 1--38, 2015.
\newblock URL \url{https://acta.astrouw.edu.pl/Vol65/n1/pdf/pap_65_1_1.pdf}.

\bibitem[Villar et~al.(2019)Villar, Berger, Miller, Chornock, Rest, Jones, Drout, Foley, Kirshner, Lunnan, et~al.]{villar2019supernova}
Villar, V., Berger, E., Miller, G., Chornock, R., Rest, A., Jones, D., Drout, M., Foley, R., Kirshner, R., Lunnan, R., et~al.
\newblock Supernova photometric classification pipelines trained on spectroscopically classified supernovae from the pan-starrs1 medium-deep survey.
\newblock \emph{The Astrophysical Journal}, 884\penalty0 (1):\penalty0 83, 2019.

\bibitem[Villar et~al.(2020)Villar, Hosseinzadeh, Berger, Ntampaka, Jones, Challis, Chornock, Drout, Foley, Kirshner, et~al.]{villar2020superraenn}
Villar, V.~A., Hosseinzadeh, G., Berger, E., Ntampaka, M., Jones, D.~O., Challis, P., Chornock, R., Drout, M.~R., Foley, R.~J., Kirshner, R.~P., et~al.
\newblock Superraenn: a semisupervised supernova photometric classification pipeline trained on pan-starrs1 medium-deep survey supernovae.
\newblock \emph{The Astrophysical Journal}, 905\penalty0 (2):\penalty0 94, 2020.

\bibitem[Wang et~al.(2021)Wang, Zhu, Adam, Yuille, and Chen]{wang2021max}
Wang, H., Zhu, Y., Adam, H., Yuille, A., and Chen, L.-C.
\newblock Max-deeplab: End-to-end panoptic segmentation with mask transformers.
\newblock In \emph{Proceedings of the IEEE/CVF conference on computer vision and pattern recognition}, pp.\  5463--5474, 2021.

\bibitem[Zhang et~al.(2024)Zhang, Helfer, Gagliano, Mishra-Sharma, and Villar]{zhang2024maven}
Zhang, G., Helfer, T., Gagliano, A.~T., Mishra-Sharma, S., and Villar, V.~A.
\newblock Maven: a multimodal foundation model for supernova science.
\newblock \emph{Machine Learning: Science and Technology}, 5\penalty0 (4):\penalty0 045069, 2024.

\end{thebibliography}
\bibliographystyle{icml2025}

\newpage
\appendix
\onecolumn

\section{SwinV2 Architecture Details}
\label{appendix:swinv2}

\begin{figure}[h]
\vskip 0.2in
\begin{center}
\centerline{\includegraphics[width=8cm]{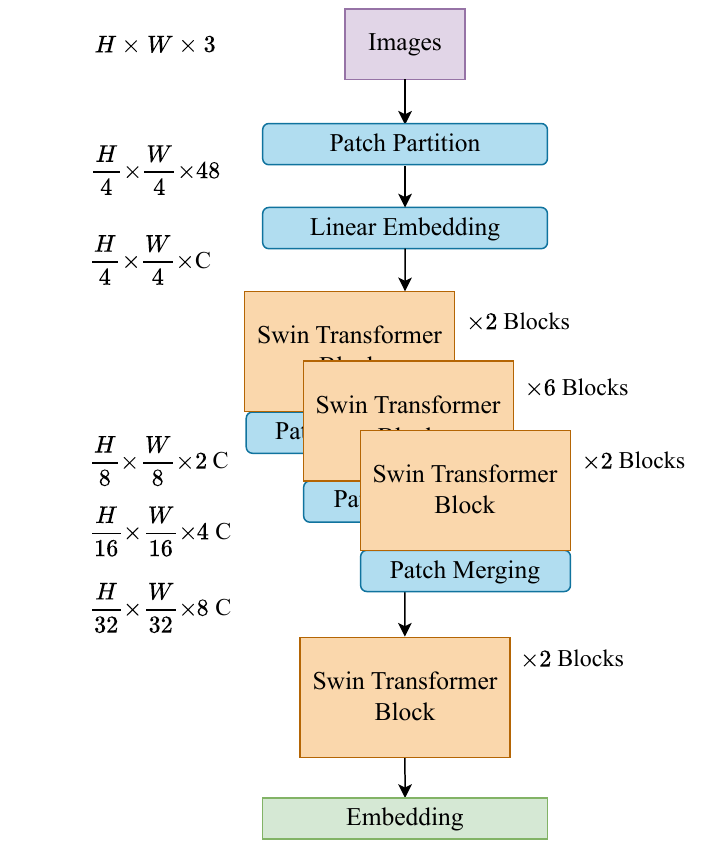}}
\caption{\textsc{SwinV2} architecture. The rounded light blue rectangles highlight the components where the model changes the dimensionality of the information, while the regular rectangles indicate where it remains fixed. The figure is from \citet{moreno2025leveraging}.}
\label{fig:swinv2_arch}
\end{center}
\vskip -0.4in
\end{figure}

The processing pipeline, illustrated in Figure~\ref{fig:swinv2_arch}, proceeds as follows. The \textsc{SwinV2} architecture first partitions each input image (Section~\ref{sec:lc_to_images}) into fixed-size patches (Patch Partition). Each patch is then flattened and projected into a higher-dimensional space via a linear transformation (Linear Embedding), producing tokens that serve as input representations for the first \textsc{Swin Transformer} Block, where Multi-head Self-Attention (MSA) is applied. These tokens first pass through Window-based MSA (W-MSA), where attention is restricted to non-overlapping local windows, reducing computational complexity from quadratic to linear with respect to the number of tokens, while capturing local dependencies. To enable cross-window interactions, the model alternates W-MSA with Shifted Window MSA (SW-MSA), which shifts the window partitions so that tokens at the edges of one window are included in neighboring windows, enabling the capture of global context. \textsc{SwinV2} builds hierarchical feature representations by progressively applying patch merging layers between \textsc{Swin Transformer} Blocks (Patch Merging), reducing spatial resolution while increasing embedding dimensionality and enabling multi-scale feature extraction. Within this framework, scaled cosine attention normalizes attention scores, improving training stability, while residual post-normalization mitigates the accumulation of large activations in deeper layers. Additionally, a log-spaced continuous position bias emphasizes local token interactions while preserving the ability to model long-range dependencies, enhancing generalization across varying image resolutions. Together, these components enable \textsc{SwinV2} to efficiently and scalably process high-resolution light curve images. After the hierarchical stages, the output of the last \textsc{Swin Transformer} Block is passed through an adaptive average pooling layer, which computes the mean across all tokens, producing a single pooled output vector that serves as the final embedding of the image. Finally, fine-tuning adapts the model to capture class-discriminative patterns for light curve classification. Further details of this approach can be found in \citet{liu2021swin} and \citet{moreno2025leveraging}.

\section{Hyperparameter Tuning for the BRF Classifier}
\label{appendix:hyperparameters}

The BRF classifier hyperparameters were optimized using F1-score on the validation set across five stratified folds, matching those used to train, validate, and test the \textsc{SwinV2} embeddings. A total of 120 combinations were evaluated via grid search over the following parameter grid: \texttt{n\_estimators} = \{100, 200, 300, 500\}, \texttt{max\_depth} = \{None, 10, 20, 30, 50\}, \texttt{max\_features} = \{'sqrt', 'log2'\}, and \texttt{min\_samples\_split} = \{2, 5, 10\}. Table~\ref{table:rf_hyperparam_tuning} reports the top 20 configurations, with F1-score summarized as mean $\pm$ standard deviation across the five validation folds.

\begin{table}[h]
\caption{Top 20 hyperparameter configurations for the BRF classifier on the validation set.}
\label{table:rf_hyperparam_tuning}
\vskip 0.15in
\begin{center}
\begin{small}
\begin{sc}
\begin{tabular}{c c c c c}
\toprule
N\_estimators & Max Features & Max Depth & Min Samples Split & F1-score \\
\midrule
500 & log2 & 30    & 5 & 65.7 $\pm$ 0.5 \\
500 & log2 & None  & 5 & 65.7 $\pm$ 0.5 \\
500 & log2 & 50    & 5 & 65.7 $\pm$ 0.5 \\
500 & log2 & 10    & 2 & 65.7 $\pm$ 1.2 \\
500 & log2 & 20    & 5 & 65.7 $\pm$ 0.5 \\
300 & log2 & 20    & 2 & 65.7 $\pm$ 1.3 \\
300 & log2 & None  & 2 & 65.6 $\pm$ 1.5 \\
300 & log2 & 30    & 2 & 65.6 $\pm$ 1.5 \\
300 & log2 & 50    & 2 & 65.6 $\pm$ 1.5 \\
500 & log2 & None  & 2 & 65.6 $\pm$ 1.2 \\
500 & log2 & 50    & 2 & 65.6 $\pm$ 1.2 \\
500 & log2 & 30    & 2 & 65.6 $\pm$ 1.2 \\
500 & sqrt & 20    & 2 & 65.5 $\pm$ 0.2 \\
500 & log2 & 20    & 2 & 65.5 $\pm$ 1.2 \\
300 & sqrt & 20    & 2 & 65.5 $\pm$ 0.3 \\
500 & sqrt & 10    & 5 & 65.5 $\pm$ 0.6 \\
500 & sqrt & 30    & 2 & 65.4 $\pm$ 0.4 \\
500 & sqrt & 50    & 2 & 65.4 $\pm$ 0.4 \\
500 & sqrt & None  & 2 & 65.4 $\pm$ 0.4 \\
500 & sqrt & 20    & 5 & 65.4 $\pm$ 0.5 \\
\bottomrule
\end{tabular}
\end{sc}
\end{small}
\end{center}
\vskip -0.1in
\end{table}

\section{Confusion Matrices of Selected Models}\label{appendix:cm_matrices}

\begin{figure*}[h]
\vskip 0.15in
\begin{center}
\centerline{\includegraphics[width=17cm]{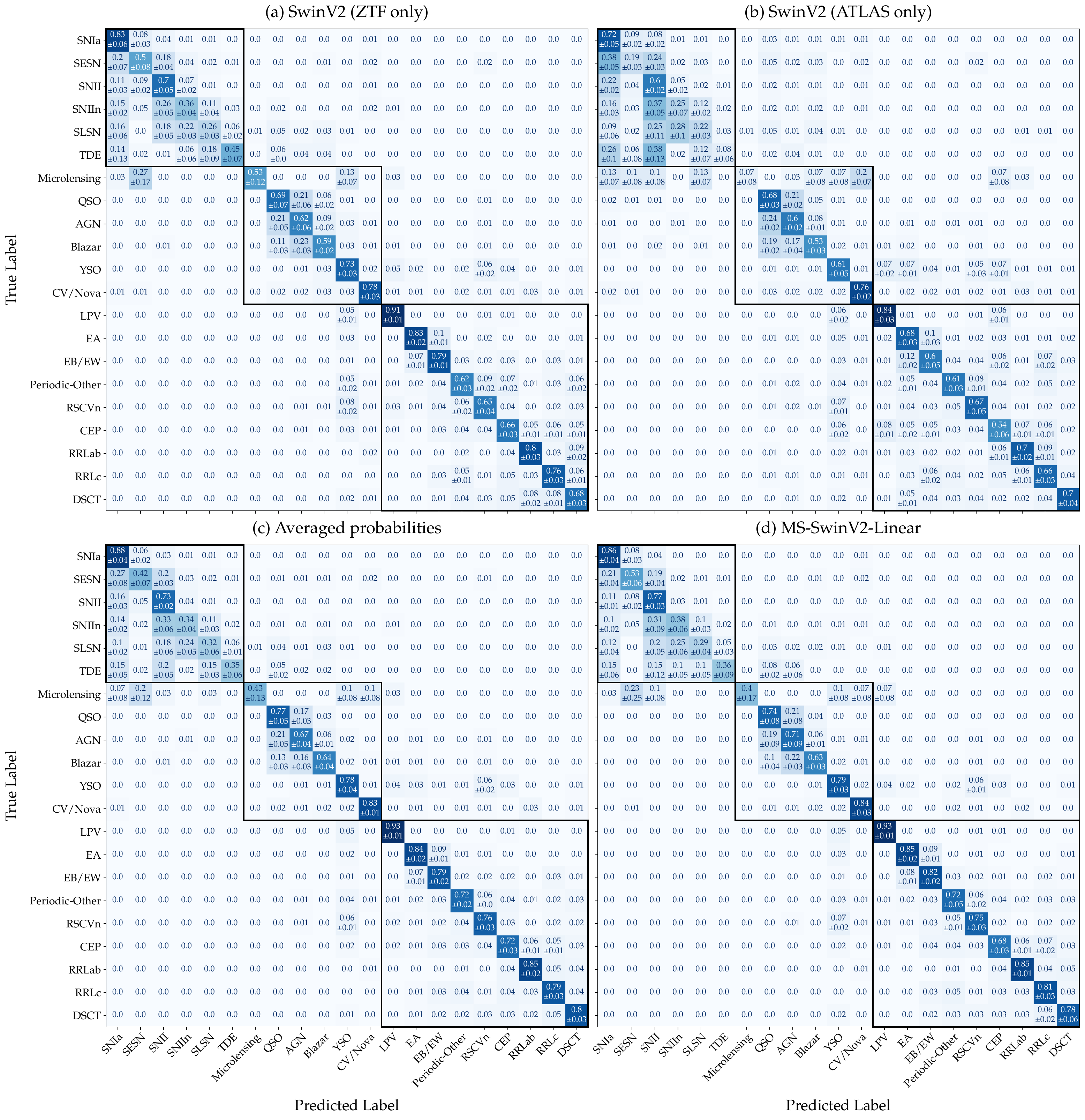}}
\caption{
Confusion matrices for the different strategies explored to combine ZTF and ATLAS data for photometric classification. 
\textbf{(a)} Single-survey \textsc{SwinV2} model trained on ZTF light curves. 
\textbf{(b)} Single-survey \textsc{SwinV2} model trained on ATLAS light curves. 
\textbf{(c)} Prediction-level fusion: combined predictions obtained by averaging the output probabilities of the single-survey ZTF and ATLAS models. 
\textbf{(d)} Multi-survey \textsc{SwinV2} architecture: jointly trained on both ZTF and ATLAS light curves, with separate input streams and a final Linear Layer classifier. 
Black lines delineate the three main groups of objects defined by the ALeRCE Team: Transient, Stochastic, and Periodic groups, from left to right. 
Standard deviations across cross-validation folds are displayed for values greater than or equal to 0.05.
}
\label{fig:cm_models}
\end{center}
\vskip -0.1in
\end{figure*}

This appendix provides the confusion matrices for the key models reported in Section~\ref{sec:results}. Specifically, we include the ZTF-only and ATLAS-only single-survey baselines (Model~3 and Model~4), the post-hoc fusion with averaged probabilities (Model~3+4), and the MS-\textsc{SwinV2}-Linear (Model~6). These visualizations offer additional insight into class-wise performance across the different fusion strategies. The confusion matrices reveal differences between the single-survey baselines. The \textsc{SwinV2} model trained on ZTF (Figure~\ref{fig:cm_models}a) achieves higher recall than the ATLAS-only baseline (Figure~\ref{fig:cm_models}b) in 19 out of 21 classes, with the ATLAS-only baseline showing higher recall only for \texttt{RSCVn} and \texttt{DSCT}. The ATLAS-only baseline exhibits particularly low recall for several rare or challenging classes, including \texttt{SESN} (which contains \texttt{SNIbc} and \texttt{SNIIb}), \texttt{SLSN}, \texttt{TDE}, and \texttt{Microlensing}, with recall values below 25\%. For the remaining ATLAS-only classes, recall reaches 50\% or higher. In contrast, the ZTF-only baseline maintains recall above 25\% for all classes, with the lowest recall observed for \texttt{SNIIn} (36\%) and \texttt{SLSN} (26\%). All other ZTF-only classes achieve recall of 50\% or higher.

Averaging the predicted probabilities from the single-survey baselines (Figure~\ref{fig:cm_models}c) results in recall increases in 17 out of 21 classes compared to the ZTF-only baseline. Recall decreases were observed in four classes where the ATLAS-only baseline recall was already below 25\%: \texttt{SESN}, \texttt{SLSN}, \texttt{TDE}, and \texttt{Microlensing}. For these, the averaged-probabilities fusion yielded lower recall than the ZTF-only baseline. The \texttt{EB/EW} class maintained the same recall as in the ZTF-only baseline. Overall, the averaged-probabilities fusion appears to leverage complementary information between surveys, producing more balanced class-wise performance compared to using only ZTF data.

The MS-\textsc{SwinV2}-Linear model (Figure~\ref{fig:cm_models}d) further increases class-wise recall in 19 out of 21 classes compared to the ZTF-only baseline, suggesting that the model can effectively leverage complementary information between surveys. The four most challenging classes, \texttt{SESN}, \texttt{SLSN}, \texttt{TDE}, and \texttt{Microlensing}, exhibited low recall in the ATLAS-only baseline (19\%, 25\%, 8\%, and 7\%, respectively), which limits the contribution of complementary information in these cases and makes improvements over the ZTF-only baseline more difficult. While the averaged-probabilities fusion did not improve recall in any of these classes, MS-\textsc{SwinV2}-Linear achieved a +3 percentage point improvement over the ZTF-only baseline in two of them: \texttt{SESN} and \texttt{SLSN}, while no improvement was observed for \texttt{TDE} and \texttt{Microlensing}, which are also the minority classes in the dataset. This suggests that joint multi-survey learning can extract useful features even when one survey contributes limited information.

\end{document}